\documentclass[fleqn,twoside]{article}
\usepackage{espcrc2}

\newcommand{\AmS}{{\protect\the\textfont2
  A\kern-.1667em\lower.5ex\hbox{M}\kern-.125emS}}
\def\Journal#1#2#3#4{{#1} {\bf #2} (#4) #3}

\def\NPB{\em Nucl. Phys.}
\def\PLB{\em Phys. Lett.}

\def\CMP{\em Commun. Math. Phys.}

\title {Finite Volume Effects in Weak Hadronic Decays \thanks{Talk 
presented by M. Testa at the Workshop on Lattice Hadron Physics 
(LHP2001), Cairns, Australia}}

\author{C.-J.D.~Lin\address[SH]{Dept.  of Physics and Astronomy, 
Univ.  of Southampton,\\
Southampton, SO17 1BJ, UK}, G.~Martinelli\address[RU]{Dip.  di Fisica, 
Univ.  di Roma ``La Sapienza'' and INFN, Sezione di Roma,\\
Piazzale Aldo Moro 2, I-00185 Rome, Italy}, 
C.T.~Sachrajda\addressmark[SH],M.~Testa\addressmark[RU].}
       
\begin{document}

\begin{abstract}
In this talk we discuss finite-volume computations of two-body 
hadronic decays below the inelastic threshold (e.g.  $K\to\pi\pi$ 
decays). In particular we show how the relation between 
finite-volume matrix elements and physical amplitudes, recently 
derived by Lellouch and L\"uscher, can be extended to all elastic states 
under the inelastic threshold. We also provide a derivation of the 
L\"uscher quantization condition directly in quantum field theory.
\end{abstract}

\maketitle

\section{Introduction}

Lattice QCD offers a natural opportunity to compute the matrix elements 
for $K\to\pi\pi$ decays from first principles.  The main difficulties 
are related to the continuum limit of the regularized theory (the 
\textit{ultra-violet} problem) and to the relation between matrix 
elements computed in a finite Euclidean space-time volume and the 
corresponding physical amplitudes (the \textit{infrared} problem).  
The ultra-violet problem, which deals with the construction of finite 
matrix elements of renormalized operators from the bare lattice ones, 
has been addressed in a series of papers~\cite{renorm}-\cite{dawson} 
and we will not consider it further.  The infrared problem arises from 
two sources:

\begin{itemize}
\item the unavoidable continuation of the theory to
Euclidean space-time and
\item the  use of a finite volume in numerical simulations.
\end{itemize}

One of the main obstacles in the extraction of physical amplitudes 
from lattice simulations stems from the rescattering of final state 
particles in Euclidean space.  The formalization of this problem, in 
the infinite-volume case, was considered in ref.~\cite{mt}.  An 
important step towards the solution of the infrared problem has 
recently been achieved by Lellouch and L\"uscher~\cite{lll} (LL), who 
derived a relation, given below in eq.(\ref{LLL}), between the 
$K\to\pi\pi$ matrix elements in a finite volume and the physical 
kaon-decay amplitudes.  For technical reasons the LL formula has been 
derived for a finite, fixed number of pion states under the inelastic 
threshold and for matrix elements at zero four-momentum transfer.  In 
the following sections a different approach to the LL formula will be 
discussed~\cite{lmst}, which extends it to all elastic states under the 
inelastic threshold and momentum transfers different from zero.

\section{Physics in a Finite Cubic Volume} \label{fvolume}

In this section we will recall some aspects of the quantization in a 
cubic box, relevant in the discussion of the LL 
formula~\cite{ml},\cite{lmst}.  In particular we will be interested in 
the structure of the zero-momentum, finite volume, energy eigenstates, 
$\left| {\pi \pi ,n} \right\rangle _V$, which can be excited from the 
vacuum by a scalar operator $\sigma (x)$ i.e.  for which 
\begin{equation}_V\left\langle {\pi \pi ,n} \right|\sigma (0)\left| 0 
\right\rangle \neq 0 \label{scalar}\ .
\end{equation}
When analyzed from the point of view of angular momentum, 
energy-eigenstates in a cubic box are a complicated 
superposition~\cite{ml}
\begin{equation}
\left| {\pi \pi ,n} \right\rangle _V=\sum_{l=0}^{\infty} 
\sum_{m=-l}^{+l}{\alpha _{l,\,m}^{(n)}\left| {\pi \pi ,n;l,m} 
\right\rangle _V}\ .
\label{expansion}
\end{equation}
We are interested in the structure of states which are cubically 
invariant {\em and} contain an s-wave component (and hence satisfy 
eq.(\ref{scalar})).

In the framework of quantum mechanics, finite-volume quantization 
formulas become exact for finite-range potentials and in the presence 
of an angular momentum cut-off, which, in its simplest form, assumes 
that only s-waves interact.  Under these conditions, the allowed 
values of the ÒradialÓ relative momentum $k$ of a two particle state, 
related to the center of mass energy $E$ as $E=2\sqrt {m_\pi ^2+k^2}$, 
are quantized as follows~\cite{ml}:
\begin{enumerate}
\item $k$ obeys the equation (see sec.\ref{sec:inelastic})
\begin{equation}
h(k,L) \equiv \frac {\phi (q) + \delta(k)}{\pi}=n \label{phase}
\end{equation}
where $n$ is a non-negative integer, $\delta (k)$ is the infinite 
volume s-wave phase-shift, $q\equiv {{kL} \over {2\pi }}$ and
\begin{eqnarray}
& & \tan \phi (q)=- \frac{\pi^{3/2} q}{Z_{00}(1;q^2)}\label{unoo}\\
& & Z_{00}(s;q^2)=\frac{1}{\sqrt {4\pi }}\sum\limits_{\underline n\in 
Z^3} {(\underline n^2-q^2)^{-s}}\ .
\end{eqnarray} \label{uno1}
\item $k^2=\underline p_n^2=\left( {{{2\pi } \over L}} 
\right)^2\underline n^2$ (free spectrum) if at least two $\underline 
p_n$ and $\underline {p}'_n$ exist with ${{\underline 
p} '_n} ^2=\underline p_n^2$ not related by a cubic transformation.  
These states are non physical and their existence is a consequence of 
the angular momentum cut-off. \label{due2}
\end{enumerate}

States of type \ref{uno1} have a non-zero s-wave component~\cite{lmst}, 
$\Psi _{E_n}^{V_{s-wave}}(r)$, undistorted by the presence of the 
boundary compared to the infinite volume s-wave function $\Psi 
_{E_n}^{\infty _{s-wave}}(r)$.  Therefore inside the volume 
$V$,
\begin{equation}
\Psi _{E_n}^{V_{s-wave}}(r) = \frac{1}{\sqrt {c(E_n)}} \Psi 
_{E_n}^{\infty _{s-wave}}(r) \label{defc}
\end{equation}
States of type \ref{due2} cannot be simply plane-waves of the form 
$e^{i\underline p_n\cdot \underline x}$, because of the presence of 
the interaction.  However the combination $\phi (\underline x) = 
e^{i\underline p_n\cdot \underline x}-e^{i{\underline p}'_n\cdot 
\underline x}$ is a solution of the Schroedinger equation even in the 
presence of the potential, because $\phi (\underline x)$ does not 
contain an s-wave component and all other angular momenta are not 
interacting~\cite{lmst}.  The spurious states $\phi (\underline x)$ 
have a non-zero cubically-invariant projection, but do not project on 
s-wave.

In conclusion, locality and the scalar character of $\sigma (x)$ imply
that
\begin{equation}
\left\langle 0 \right|\sigma (0)\left| {\pi \pi ,n;Spurious} 
\right\rangle _V=0\ ,
\end{equation}
\begin{eqnarray}
&&\left\langle 0 \right|\sigma (0)\left| {\pi \pi ,n} \right\rangle_V 
=\int_V d^3xS(r)\Psi _{E_n}^V(\underline x) \nonumber \\
& &=\int_V d^3xS(r)\Psi _{E_n}^{V_{s-wave}}(r) = \label{prop}\\
& &=\frac{1}{\sqrt {c(E_n)}} \int d^3xS(r) \Psi 
_{E_n}^{\infty _{s-wave}}(r)\ .
\nonumber
\end{eqnarray}
where $S(r)$ is the coordinate representation of $\sigma (x)$.  $S(r)$ 
is a function of $r$ only, so that it selects the s-wave component 
$\Psi _{E_n}^{V_{s-wave}}(r)$ in the angular momentum expansion, 
eq.(\ref{expansion}).  $S(r)$ is also localized inside $V$, thus 
justifying the last step in eq.(\ref{prop}). With the definition
\begin{eqnarray}
& & \sigma(E_n)\equiv \left\langle 0 \right|\sigma (0)\left| {\pi \pi 
,E_n}\right\rangle =  \nonumber\\
& & =\int d^3xS(r) \Psi _{E_n}^{\infty _{s-wave}}(r)
\end{eqnarray}
we therefore have
\begin{equation}
\left| {\sigma (E_n)} \right|^2=c(E_n)\left| {\left\langle 0 
\right|\sigma (0)\left| {\pi \pi ,n} \right\rangle _V} \right|^2\ .
\label{factor}
\end{equation}
After these preliminaries we are ready to discuss the LL proposal, 
which consists in tuning the volume $V$ so that one of the first 
seven excited two-pion state is degenerate in energy with the 
kaon state (for $n=1$, $L\approx 5\div 6\;$Fm) and then using the LL 
relation which connects finite and infinite volume matrix elements
\begin{eqnarray}
& & \left| {\left\langle {\pi \pi ,E=m_K} \right|{\cal H}_W(0)\left| 
K \right\rangle } \right|^2=\label{LLL}\\
& & =V^2\left| {{}_V\left\langle {\pi \pi ,E} \right|{\cal H}_W(0)\left| K 
\right\rangle _V} \right|^2\left( {{{m_K} \over {k}}} 
\right)^3 \times \nonumber \\
& &	\times 8\pi [q\phi '(q)+k\delta '(k)]\ . \nonumber
\end{eqnarray}
In eq.(\ref{LLL}) $\left| {\pi \pi ,E} \right\rangle _V$ denotes a 
finite volume two pion state with zero total momentum and ``angular 
momentum'', whose energy $E$ is to be chosen among the solutions of 
eq.(\ref{phase}), while $\left| K \right\rangle _V$ denotes a finite 
volume kaon state with zero momentum.  Both states are normalized to 
$1$.  The LL formula has been derived~\cite{lll} for a large enough 
volume, $n=1\div 7$ and $\Delta E=\Delta\underline {P}=0$.

Eq.(\ref{factor}) shows that $c(E_n)$, defined in eq.(\ref{defc}), is 
precisely the LL proportionality factor.

\section{The Nature of the LL Relation}

In order to relate the states at finite and infinite volume we 
consider the correlator
\begin{eqnarray}
&&\int\limits_V {d^3x\left\langle {\sigma (\underline 
	x,t)\sigma (0)} \right\rangle 
	_V}\mathrel{\mathop{\kern0pt\longrightarrow}\limits_{V\to \infty 
	}}\label{uno} \\
&&{{(2\pi )^3} \over {2(2\pi )^6}}\int {{{d\underline p_1} \over 
	{2\omega _1}}{{d\underline p_2} \over {2\omega _2}}\delta (\underline 
	p_1+\underline p_2)e^{-(\omega _1+\omega _2)t}}\times\nonumber \\
& & \times | \langle 0 
	|\sigma (0) | \underline p_1,\underline p_2 \rangle 
	|^2=\nonumber\\
	& &={1 \over {2(2\pi )^3}}\int {dE}e^{-Et}\left| {\left\langle 0 
	\right|\sigma (0)\left| {\pi \pi ,E} \right\rangle } 
	\right|^2 \times \nonumber\\
	& & \times \int 
	{{{d\underline p_1} \over {2\omega _1}}{{d\underline p_2} \over 
	{2\omega _2}}}\delta (\underline p_1+\underline p_2)\delta (E-\omega 
	_1-\omega _2)=\nonumber\\
	& &\hspace{-0.2in}
={\pi  \over {2(2\pi )^3}}\int {{{dE} \over E}}e^{-Et}\left| 
	{\left\langle 0 \right|\sigma (0)\left| {\pi \pi ,E} \right\rangle } 
	\right|^2k(E). \nonumber
\end{eqnarray}
On the other hand we could proceed differently
\begin{eqnarray}
	& &\int\limits_V {d^3x\left\langle {\sigma (\underline 
	x,t)\sigma (0)} \right\rangle }= \label{due} \\
& & V\sum\limits_n {\left| {\left\langle 
	0 \right|\sigma (0)\left| {\pi \pi ,n} \right\rangle _V} 
	\right|^2e^{-E_nt}}\mathrel{\mathop{\kern0pt\longrightarrow}\limits_{V\to 
	\infty }}\nonumber\\
	& &\hspace{-0.2in}
\mathrel{\mathop{\kern0pt\longrightarrow}\limits_{V\to \infty 
	}}V\int\limits_0^\infty  {dE\rho_{V}(E)\left| {\left\langle 0 
	\right|\sigma (0)\left| {\pi \pi ,E} \right\rangle _V} 
	\right|^2e^{-Et}}\,, \nonumber
\end{eqnarray}
where only states containing an s-wave 
component contribute.  In eq.(\ref{due}) $\rho_V(E)$ denotes a function to be 
determined, which provides the correspondence between finite volume 
sums and infinite volume integrals.  In many cases, for example in one 
dimension, $\rho_V(E)$ can be identified as the density of states at 
energy $E$.  In section \ref{sec:large} we show that, also in 
the presence of cubic boundary conditions, $\rho _{V}(E)$ is given by
\begin{equation}
\rho _{V}(E)=\frac{dn}{dE} = \frac{q\phi^\prime(q)
+k\delta^\prime(k)}{4 \pi k^2}E\ ,\label{density}
\end{equation}
with exponential precision in the volume.  The expression in 
eq.~(\ref{density}) is the one one would heuristically derive from a 
na\"{\i}ve interpretation of $\rho_V(E)$ as the density of states, as 
seen from eq.~(\ref{phase}).

Comparing eqs.({\ref{uno}) and ({\ref{due}) we get the correspondence 
\begin{equation}
\left| {\pi \pi ,E} \right\rangle \Leftrightarrow 4\pi \sqrt 
{{{VE\rho_{V} (E)} \over {k(E)}}}\left| {\pi \pi ,E} \right\rangle _V\,.
\end{equation}
In a similar way one gets
\begin{equation}
\left| K, {\underline p=0} \right\rangle \Leftrightarrow 
\sqrt {2mV}\left|K, {\underline p=0} \right\rangle _V\,,
\end{equation}
from which it is easy to recover the LL relation, eq.(\ref{LLL}), 
without any restriction on the four-momentum transfer.

Although the present approach appears superficially to be equivalent 
to the one of ref.~\cite{lll}, there is an important difference in the 
two derivations.  The result of ref.~\cite{lll} was obtained at a 
fixed value of $n$ and therefore at a fixed volume $V$, tuned so that 
$m_K=E_{n}$, with $n < 8$.  We, on the other hand, derived the same 
result at fixed energy $E$, for asymptotically large volumes 
$V$.  This implies that, as $V \rightarrow \infty$, we must 
simultaneously allow $n \rightarrow \infty$.  A question which arises 
naturally at this point is what is the relation between the two 
approaches?  The answer requires a more detailed discussion, developed 
in the following section, where it will be shown that the constraints 
of locality allow us to establish eq.~(\ref{density}) with exponential 
accuracy for elastic states under the inelastic threshold.

\section{Summation Theorems, Locality and the LL 
Formula}\label{sec:large}

Locality has already been an important ingredient in establishing 
eq.(\ref{factor}).  In this section we will discuss another important 
and more subtle r\^ole of locality.  Our approach to the LL formula, 
outlined in the previous section, is based on the identification
\begin{eqnarray}
&&\sum\limits_n {\left| {\left\langle 0 \right|\sigma (0)\left| {\pi \pi 
,n} \right\rangle _V} \right|^2e^{-E_nt}}\approx\nonumber\\ 
&&\approx \int\limits_{\bar 
E}^\infty {dE\left|\,\sigma(E) \right|^2e^{-Et}}
\end{eqnarray}
of finite and infinite volume correlators.  We are therefore naturally 
led to the question of how well a sum may approximate an integral.  We 
will show that the key ingredient to answer this question is again 
locality.

We start with the example of a simple summation theorem.  Take any 
$\tilde f(\underline x)$ and $\tilde g(\underline x)$ exponentially 
decreasing at large $|\underline x |$ and compute their Fourier transforms
\begin{eqnarray}
f(\underline p)=\int\limits_{\scriptstyle V\hfill\atop
  \scriptstyle \infty \hfill} {d\underline x\;\tilde 
f(\underline x)e^{i\underline p\cdot \underline x}}\,\, ,
\, g(\underline p)=\int\limits_{\scriptstyle V\hfill\atop
  \scriptstyle \infty \hfill} {d\underline x\;\tilde 
g(\underline x)e^{i\underline p\cdot \underline x}}.
\label{ft1}
\end{eqnarray}
In eq.(\ref{ft1}) the integrals can be taken on a finite volume $V$ or 
over all the three-dimensional space, the difference being 
exponentially small in $V$.  If we define
\begin{equation}
\underline p_n\equiv \frac{2 \pi}{ L}\underline n
\label{spectrum}
\end{equation}
we have
\begin{eqnarray}
& &\hspace{-0.25in} \int_\infty \tilde f(\underline x)\tilde
g^{\ast}(\underline x) d\underline x=\frac {1}{(2\pi )^3}\int
f(\underline p)g^{\ast}(\underline p)d\underline p = \nonumber \\ & &
\hspace{-0.25in}=\int_V \tilde f(\underline x)\tilde
g^{\ast}(\underline x) d\underline x=\frac{1}{V}\sum_{\{ \underline
p_n\}} f(\underline p_n)g^*(\underline p_n)
\end{eqnarray}
and therefore
\begin{equation}
\frac{1}{(2\pi )^3}\int f(\underline p)g^*(\underline 
p)d\underline p= \frac{1}{V}\sum_{\{\underline p_n\}} 
f(\underline p_n)g^*(\underline p_n)\,,
\label{summation}
\end{equation}
again with exponential accuracy in $V$.  If the support of $\tilde 
f(\underline x)$ and $\tilde g(\underline x)$ is entirely contained 
inside $V$, eq.(\ref{summation}) becomes exact.

Relations similar in nature to eq.(\ref{summation}) can be obtained in 
quantum mechanics and field theory.  In fact, since $\sigma 
(x)$ is a local operator, $\sigma (0)\left| 0 \right\rangle$ is a 
localized state which does not differ much from the vacuum state away 
from $0$.  This is a consequence of clustering, which guarantees that, 
in the absence of massless particles, the probability of finding 
particles at a distance $r$ away from the origin in the state $\sigma 
(0)\left| 0 \right\rangle$ decreases exponentially like $e^{-2m_{\pi} 
r}$.  As a consequence, if $V$ is greater that the 
localization volume, we can write, at $t=0$,
\begin{eqnarray}
\lefteqn{\int_V d^3x \langle \sigma (\underline x,0)\sigma 
(0) \rangle_V =V \sum_n | \langle 0 |\sigma (0) | \pi \pi ,n \rangle _V |^2}
\nonumber\\  
& = &\int_{\bar E}^\infty dE | \sigma (E) |^2 = \int d^3x \langle 
\sigma (\underline x,0)\sigma (0) \rangle
\label{complete}\end{eqnarray}
with exponential accuracy in $V$.  The second and third lines of
eqs.(\ref{complete}) were obtained by inserting the complete set of
energy eigenstates on a finite and infinite volume respectively.  The
result is the same apart from the exponentially small perturbation at the 
boundary.  In other words, we can introduce states
\begin{equation}
\left| {\pi \pi ,\underline x} \right\rangle _V=\sqrt 
V\sum_n \left| \pi \pi ,n \right\rangle _V\,{\Psi 
_{E_n}^{V}}^{\!\ast}(\underline x)
\end{equation}
where $\Psi_{E_n}^{V}(\underline x)$ are the finite volume center of 
mass wave functions.  $\left| {\pi \pi ,\underline x} \right\rangle 
_V$ represents a state with $\underline P=0$, in which the two pions 
are localized a distance $\underline x$ apart.  If the distance 
$\underline x$ is inside $V$, the same state is exponentially 
well represented by
\begin{equation}
\left| {\pi \pi ,\underline x} \right\rangle= (2 \pi )^{3/2} 
\int_{\bar E}^{\infty} dE \left| \pi \pi ,E \right\rangle {\Psi 
_E^{\infty}}^*(\underline x)
\end{equation}
where $\Psi _E^{\infty}(\underline x)$ are the infinite-volume 
center of mass wave functions.

By virtue of the cluster property mentioned above, we know that
\begin{equation}
F(\underline x)\equiv \left\langle \pi \pi ,\underline x \right|\sigma 
(0)\left| 0 \right\rangle
\end{equation}
is a localized function of $\underline x$, so that
\begin{equation}
F(\underline x)\approx {}_{V}\langle \pi \pi ,\underline x |\sigma (0) | 
0 \rangle
\end{equation}
with exponential precision and
\begin{eqnarray}
& &\int_V d^3x \langle \sigma (\underline x,0)\sigma (0) \rangle_{V} = 
\nonumber \\
& & =V\sum_n | \langle 0 |\sigma (0) | \pi \pi ,n \rangle _V |^2 =\nonumber \\
& & = \int | \langle 0 |\sigma (0) | \pi \pi ,\underline x \rangle_{V} 
|^2 d\underline x= \\
& & = \int | \langle 0 |\sigma (0) | \pi \pi ,\underline x 
\rangle |^2 d\underline x =\nonumber \\ 
& & =\int_{\bar E}^\infty  dE | \sigma (E)
|^2=\int d^3x \langle \sigma (\underline x,0)\sigma 
(0) \rangle\nonumber 
\end{eqnarray}
confirming eq.(\ref{complete}).  Eqs.(\ref{complete}) and (\ref{defc}) 
express the important property that the matrix elements of a local 
operator are smooth in energy, in the sense that the sum over discrete 
energy levels reproduces, with exponential accuracy, the corresponding 
integral
\begin{eqnarray}
\int_{\bar E}^\infty dE | \sigma (E) |^2= V \sum_{n}\frac{| \sigma 
(E_n) |^2}{c(E_n)}\ .
\end{eqnarray}

Introduction of time dependence does not substantially change this 
result.  We have
\begin{eqnarray}
& &\int_V d^3x \langle \sigma (\underline x,t)\sigma 
(0) \rangle _{V} =\label{time} \\
& &=V\sum_n | \langle 0 |\sigma (0) | \pi 
\pi ,n \rangle _V |^2e^{-E_n t}=\nonumber \\
& &=\int \langle 0 |\sigma (0) | \pi \pi ,\underline x
\rangle e^{-\tilde Ht} \langle \pi \pi ,\underline x |\sigma 
(0) | 0 \rangle d\underline x=\nonumber \\
& &\hspace{-0.25in}
=\int F^*(\underline x)(e^{-\tilde Ht}F )(\underline x)d\underline 
x=
\int  F^*(\underline 
x)F(\underline x,t)d\underline x, \nonumber 
\end{eqnarray}
where the two-body hamiltonian $\tilde H$ is related to the second 
quantized hamiltonian, $H$, through
\begin{eqnarray}
\lefteqn{e^{-Ht}\left| {\pi \pi ,\underline x} \right\rangle 
=\sqrt V\sum_n | \pi \pi ,n \rangle _V\Psi 
_{E_n}^*(\underline x)e^{-E_nt}=}\nonumber \\
& &=\sqrt V\sum_n | \pi \pi ,n \rangle 
_Ve^{-\tilde Ht}\Psi _{E_n}^*(\underline x)\ .
\end{eqnarray}
$F(\underline x,t)$ as defined in eq.(\ref{time}) has a time evolution 
characteristic of a diffusion process with hamiltonian $\tilde H$.  
Therefore both $F(\underline x,t)$ and the correlator $\int_V d^3x 
\langle \sigma (\underline x,t)\sigma (0) \rangle_{V}$ are insensitive 
to the presence of the boundaries, up to exponentially small terms, 
for a long time ($t\approx m_{\pi} L$), and we have
\begin{eqnarray}
\int_{\bar E}^\infty dE |\,\sigma (E) |^2 e^{-Et}=\label{prepoisson}
V\sum_n \frac{| \sigma(E_{n})|^2}{c(E_{n})}e^{-E_nt}
\end{eqnarray}
with exponential precision in $V$.

The basic tool for relating integrals to sums is provided by the Poisson 
identity
\begin{equation}
\sum\limits_{n=-\infty }^{+\infty } {\delta 
(n- x)}=\sum\limits_{l=-\infty }^{+\infty } {e^{2\pi ilx}}\,.
\label{identity}
\end{equation}
Eq.(\ref{identity}), together with the substitution $x \to h(E,L)$, 
where $h(E,L)$ is defined in eq.(\ref{phase}), gives
\begin{eqnarray}
& &\int\limits_{\overline 
E}^\infty  {dE\left| {\sigma (E)} \right|^2e^{-Et}}=\label{poisson1}\\
& &=\sum\limits_n {{{\left| {\sigma (E_n)} \right|^2e^{-E_nt}} \over 
{\left. {{{\partial h(E,L)} \over {\partial E}}} 
\right|_{E_n}}}- {\cal Q} (L,t)}\,, \nonumber
\end{eqnarray}
where
\begin{equation}
{\cal Q}(L,t)\equiv \sum_{l\ne 0} 
\int_{\bar E}^\infty dE | \sigma (E)
|^2 e^{-Et} e^{2il \delta (k)} e^{2il \phi (q)}.
\label{qdef}
\end{equation}
Consider the large $L$ behavior of ${\cal Q}(L,t)$.  The 
techniques of asymptotic analysis~\cite{bleis}, together with the fact 
that $\phi(q)\to\infty$ when $L\to\infty$ at fixed $k$,
suggest that the large $L$ behavior of each term of the sum over $l$ 
in eq.(\ref{qdef}) is dominated by the critical points of $\phi (q)$, 
i.e.  the points at which $\phi (q)$ has a vanishing derivative or the 
points where $\phi (q)$ and $\sigma (E)$ are not continuous or 
differentiable.  $q=0$ is a critical point of $\phi (q)$ as follows 
from
\begin{equation}
\phi (q)\mathop \approx \limits_{q\approx 
0}2\pi ^2q^3
\end{equation}
so that, in absence of further critical points we expect that
\begin{equation}
Q(L,t)\mathrel{\mathop{\kern0pt\longrightarrow}\limits_{L\to \infty 
}}0
\end{equation}
as a power of ${1 \over L}$ depending on the threshold behavior of 
$\sigma (E)$ which, for local observables, may be make arbitrarily 
small~\cite{lmst}.

A comparison between eqs.(\ref{prepoisson}) and (\ref{poisson1}) then 
leads to the identification
\begin{equation}
\left.  {c(E_n)\Leftrightarrow V{{\partial 
h(E,L)} \over {\partial E}}} \right|_{E_n}
\end{equation}
and gives the LL relation with exponential accuracy, for energies 
between the elastic and inelastic thresholds.

Although a rigorous mathematical proof that $q=0$ is the only critical 
point of $\phi (q)$ in the complex $q$-plane is lacking, we can 
provide a simple, indirect argument that this is likely to be the case.  For 
this we start applying eq.(\ref{summation}) to the case of two 
functions $f(p)$ and $g(p)$ which only depend on $p \equiv |\underline 
p|$
\begin{eqnarray}
\lefteqn{\frac{1}{2\pi^2}\int_{0}^{\infty} f( p)g^*(p) p^{2} d p= 
\frac{1}{V}\sum_{\{\underline p_n\}} f(p_n)g^*( p_n)=}\nonumber\\
& & = \frac{1}{V}\sum_{\{ p_n\}} \nu_{n} f(p_n)g^*( p_n)\,,
\label{summationp}
\end{eqnarray}
where $\nu _n$ is the number of integer vectors with given $\left| 
{\underline n} \right|$.  The last sum in eq.(\ref{summationp}) is 
performed over the distinct values of $p_n$.  On the other hand
the $p_{n}$'s are the solutions of
\begin{equation}
\phi (q)=n \pi
\end{equation}
because, from eq.(\ref{unoo}) we have that
\begin{equation}
\tan \phi (q)=0\Rightarrow q=\left| {\underline n} \right|
\end{equation}
where $\underline n$ is any vector with integer components.
We can now use the Poisson Identity eq.(\ref{identity}) with $x\to
\phi(q)/\pi$
which, together with
\begin{eqnarray}
\phi '(| \underline n |) |=\frac {4\pi ^2} 
{\nu _n} \underline n^2
\end{eqnarray}
and after multiplication by $p^2 / 2\pi ^2 f(p)g^*(p)$ and integration 
over $p$, gives
\begin{eqnarray}
\lefteqn{\frac {1}{3V} f(0)g^*(0)+\frac {1}{V}\sum_{\{p_n\ne 
0 \}} \nu _n f(p_n)g^*(p_n)=}\nonumber\\
\lefteqn{=- \frac {2}{3V} f(0)g^*(0)+\frac{1}{V}\sum_{\{ p_n \}} \nu _n 
f(p_n)g^*(p_n)=} \nonumber\\
& =&\frac {1}{(2\pi )^3}\int f(p)g^*(p)d{\underline p}+ \nonumber\\
& +&\sum_{l\ne 0} \frac {1}{2\pi ^2}\int_0^\infty
p^2f(p)g^*(p)e^{2il\phi (\frac {Lp}{2\pi})}dp.
\label{examp}
\end{eqnarray}
Eq.(\ref{examp}), together with eq.(\ref{summationp}) shows that
\begin{equation}
\sum_{l\ne 0} \frac{1}{2\pi ^2}\int_0^\infty  
p^2f(p)g^*(p)e^{2il\phi (\frac {Lp}{2\pi})}dp
\label{exp}
\end{equation}
is indeed exponentially small with power corrections concentrated at
threshold. This is in complete agreement with the heuristic analysis
presented after eq.(\ref{qdef}).  This argument strongly suggests that
the only critical point of $\phi(q)$ is at $q=0$.  In fact if
$\phi(q)$ possessed other critical points in the complex $q$-plane,
these would show up in eq.(\ref{exp}), through a large $L$ behavior
different from the one just found.

\section{Finite Volume Quantization in Quantum Field Theory and the 
Effects of Inelasticity}\label{sec:inelastic}

The approach presented in previous sections is based on the property 
that local correlators may be expressed, with exponential accuracy, 
both as a sum or as an integral over intermediate states.  This may 
appear to be a great difference of our approach compared to that of 
ref.~\cite{lll}.  We will argue, however, that the validity of 
eq.(\ref{phase}) also requires the volume to be sufficiently large for 
the Fourier series to be equal to the infinite-volume energy 
integrals, up to exponential corrections.  Since the derivation of the 
LL relation in ref.~\cite{lll} relies on this quantization formula, we 
conclude that the conditions on the volume for the applicability of 
this relation are equivalent in the two approaches.

The validity of eq.(\ref{phase}) in Relativistic Quantum Field Theory 
(RQFT) has been discussed in ref.~\cite{ml}.  In this section we 
describe a different approach which helps to clarify the size of the 
corrections to eq.(\ref{phase}) due to the presence of an inelastic 
threshold, $E_{in}$~\cite{lmst}.

The concept of a wave function in RQFT is an approximate one.  The 
object closest to a wave function is the Bethe-Salpeter (BS) wave 
function.  In infinite-volume, for an incoming state with total 
momentum zero, the $t=0$ BS wave function is defined as
\begin{eqnarray}
& & \Phi _{\underline k}(\underline x)= \langle 0 | \phi (\underline 
x,0)\phi (0) | \underline k,-\underline k \rangle _{in} = \nonumber\\
& & =\sum_n \langle 0 
|\phi (\underline x,0) | n \rangle \langle n 
|\phi (0) | \underline k,-\underline k \rangle _{in}\,,
\label{bethe}\end{eqnarray}
where $\phi (x)$ is an appropriately normalized pion field.  In 
eq.(\ref{bethe}) we can separate the contribution of single pion 
states, $\Psi _{\underline k}(\underline x)$ and that of multipion 
states, $I_{\underline k} (\underline x)$, as
\begin{equation}
\Phi _{\underline k}(\underline x)=\Psi _{\underline k}(\underline x)+ 
I_{\underline k}(\underline x) \label{additional}
\end{equation}
where
\begin{eqnarray}
&&\hspace{-0.3in}
\Psi _{\underline k}(\underline x)=\int {{{d^3 p} \over {(2\pi 
)^32E_p}}\left\langle {\underline p} \right|\phi (0)\left| {\underline 
k,-\underline k} \right\rangle _{in}e^{i\underline p\cdot \underline 
x}} \\ 
&& \hspace{-0.3in}I_{\underline k}(\underline x) = \sum_{\{3 \pi\}} \langle 0 
|\phi (0) | 3 \pi \rangle \langle 3 \pi |\phi (0) | 
\underline k,-\underline k \rangle _{in}e^{i \underline p_{3 
\pi}\cdot\underline x} \nonumber\\ 
&&\hspace{1.5in}+\cdots
\label{inelastic}
\end{eqnarray}
where the second sum runs over three pion states and the ellipses
represent the contribution from states with a higher number of
particles.  As will be shown later, $I_{\underline k}(\underline x)$
vanishes exponentially with $\underline x$ and will be neglected for
the moment.  We can further separate, in $\Psi _{\underline
k}(\underline x)$, the connected part
\begin{eqnarray}
& & \Psi _{\underline k}(\underline x)=e^{i\underline k\cdot \underline 
x}+ \label{connected}\\
& & + \int {{{d^3p} \over {(2\pi )^3 2 E_p}}\left\langle {\underline 
p} \right|\phi (0)\left| {\underline k,-\underline k} \right\rangle 
_{in}^{conn}e^{i\underline p\cdot \underline x}} \nonumber
\end{eqnarray}
and parametrize $\langle \underline p |\phi (0) | \underline 
k,-\underline k \rangle _{in}^{conn}$ as
\begin{eqnarray}
\langle \underline p |\phi (0) | \underline 
k,-\underline k \rangle _{in}^{conn}= \frac{1} 
{4E_k}\frac {\cal M} {E_p-E_k -i\varepsilon }\,. \label{offshell}
\end{eqnarray}
The off-shell scattering amplitude $\cal M$ becomes the physical one 
\footnote {As before we keep the interaction only for the s-wave.}, 
${\cal M}(\underline k\to \underline p)$, when $p=k$ {\begin{equation} 
{\cal M}(\underline k\to \underline p)={{4\pi } \over i}{{2E_k} \over 
k}(e^{i2\delta (k)}-1)\,.
\end{equation}
We therefore have
\begin{eqnarray}
& & \Psi _{\underline k}(\underline x)= e^{i\underline k\cdot 
\underline x}+\label{elastic}\\
& &  +\int \frac {d^3 p} {(2\pi )^32 E_p}\frac{\cal M} 
{4E_k(E_p-E_k-i\varepsilon )}e^{i\underline p\cdot \underline x}\,. 
\nonumber
\end{eqnarray}
The projection of $ \Psi _{\underline k}(\underline x)$ over the 
s-wave is
\begin{eqnarray}
& & \Psi _{\underline k}|_{s-wave} (r) = \frac 
{\left(e^{i2\delta (k)}+1 \right)}{2}\frac{ \sin kr} { kr}+\nonumber \\
& & +{\cal P}\int \frac {d^3 p} {(2\pi )^32E_p}\frac{(E_p+E_k)\cal M} 
{4E_k(p^{2}-k^{2})}\frac{ \sin p r} {p r}\,, \label{integral}
\end{eqnarray}
where we used the identity
\begin{equation}
\frac{1}{x - i \varepsilon}=i\pi\,\delta(x) + {\cal 
P}\frac{1}{x}\ .
\label{final}
\end{equation}
From eq.(\ref{integral}) we obtain the large $r$ behaviour of 
$\Psi _{\underline k}|_{s-wave} (r) $
\begin{equation}
\Psi _{\underline k} |_{s-wave}(r) =\frac {1}
{kr} \left( \sin kr+\frac {1} {2i} (e^{i2\delta (k)}-1) e^{ikr}
\right)\,.\label{asympt}
\end{equation}
As discussed in section \ref{fvolume}, the relative momenta $k$ 
allowed in a box are determined by the condition that the s-wave 
projection of the finite volume BS wave function, $\Phi _{\underline 
k}^{V} |_{s-wave} (r) $, is not deformed by the presence of the 
boundary.  $\Phi _{\underline k}^{V}(\underline x)$ is given by an 
expression similar to eq.(\ref{bethe}), where the sum over 
intermediate states runs over finite volume energy eigenstates, so 
that, neglecting for the moment the contribution from multipion states, 
we have
\begin{eqnarray}
& & \Psi _{\underline k}^{V} |_{s-wave} (r) \equiv \label{discrete}\\
& & \equiv \frac{1}{V} \sum_{\{\underline p_{n}\}} \frac {1} 
{2E_{p_{n}}}\frac{(E_{p_{n}}+E_k)\cal M} {4E_k 
(p_{n}^{2}-k^{2})}\frac{ \sin p_{n} r} {p_{n} r}\,, \nonumber
\end{eqnarray}
where the sum is over single particle momenta given by 
eq.(\ref{spectrum}).  The quantization condition requires therefore 
the identity of eqs.(\ref{integral}) and (\ref{discrete}).  We stress 
that in the finite volume expression, eq.(\ref{discrete}), 
disconnected terms and $i \varepsilon$'s are absent because the 
eigen-momenta $k$ do not coincide with any of the $p_{n}$.  These 
terms will appear as $V$ grows, as shown in a moment.

In appendix C of ref.~\cite{lmst} we prove the summation formula
\begin{eqnarray}
& & \frac{1}{V}{\sum_{\{\underline p_n\}}}\frac{f(p_n^2)}
{k^2-p_n^2}= \label{eq:summation}\\
& & = \frac{1}{(2\pi)^3}{\cal P}\int_{-\infty}^{\infty}d^{\,3}p\,
\frac{ f(p^2)}{k^2-p^2}\,+\,c f(k^2)
\nonumber
\end{eqnarray}
\begin{equation}
\textrm{where}\hspace{0.2in} 
c= - \frac {Z_{00}(1,q^{2})}{2 \pi ^{\frac{3}{2}} L}\ .
\label{correction}
\end{equation}
Eq.(\ref{eq:summation}) is valid up to exponentially small 
corrections, provided the Fourier transform of $f(p^2)$ vanishes 
exponentially with $r$ and, together with eqs.(\ref{discrete}) and 
(\ref{integral}), gives at once the L\"uscher quantization condition, 
eq.(\ref{phase}), with exponential precision in $V$.  The size of the 
corrections are determined by the precision with which the sum in 
eq.(\ref{discrete}) is able to reproduce the corresponding integral, a 
condition analogous to the one which was underlying our derivation of 
the LL formula.  It appears therefore that the two formulas have the 
same conditions of applicability.  In particular, a possible source of 
concern is the (practically relevant) situation in which the 
quantization volume allows only very few (perhaps two or three) 
elastic states under the inelastic threshold.  In this case, even 
though the finite-volume effects are exponentially small, one is 
working in a fixed volume and in order to establish that the 
corrections are indeed negligible one needs an estimate of such 
effects.  These corrections are small if $\Psi _{\underline k}(r) 
|_{s-wave}$ is well approximated by its asymptotic expression, 
eq.(\ref{asympt}), at $r \approx L$.  There are essentially two types 
of exponentially small corrections.  The first type is analogous to 
the corrections which would be present, even in quantum mechanics in 
the presence of an exponentially decreasing potential, rather than a 
finite range one.  The second type of corrections, characteristic of 
RQFT, are due to the existence of inelasticity and are responsible for 
the failure of eq.(\ref{phase}) above $E_{in}$.  They come from the 
term $I_{\underline k}(\underline x)$ in eq.(\ref{additional}) and a 
corresponding crossed contribution in $\Psi _{\underline k}(\underline 
x)$.  The matrix element $\langle 3 \pi |\phi (0) | \underline 
k,-\underline k \rangle _{in}$ in eq.(\ref{inelastic}) has a 
singularity when the virtual pion described by $\phi$ reaches the 
mass-shell.  This singularity is complex for $2 E_{k} < E_{in}$, 
implying an exponential vanishing of $I_{\underline k}(\underline x)$.  
For values of $2 E_{k}$ closer to $E_{in}$ the singularity is closer 
to the real axis, and the range of the exponential increases.    
Therefore the quantization condition is, in general, affected by the 
presence of inelasticity.  In the case of pions, however, there are 
indications that inelasticity is negligible up to around 1~Gev~\cite{daf}.
This shifts the singularity further from the real axis, making the 
finite-volume effects less important and allowing 
the LL strategy to be applicable to $K\to\pi\pi$ decays without
sizeable corrections.

\section{Conclusions} We have reviewed finite-volume effects in
$K\to\pi\pi$ decays, examining the conditions under which the LL
formula is valid. We were able to remove the restrictions of zero
momentum transfer and $n<8$, present in the original LL derivation,
allowing the infinite volume limit to be taken. We have also presented
an alternative simple derivation of the L\"uscher quantization
condition in Quantum Field Theory, which allows us to examine the
general consequences of inelasticity on the quantization formula. For
the particular case of $K\to\pi\pi$ decays we conclude that these
corrections are not important.

\section*{Acknowledgements}
M. Testa thanks the organizers of LHP2001 for the hospitality in
Cairns.  D. Lin, C. Sachrajda and M. Testa thank the Institute for
Nuclear Theory at the University of Washington for its hospitality and
the Department of Energy for partial support during the completion of
this work. This work was supported by European Union grant
HTRN-CT-2000-00145.

\end{document}